\documentclass[%
 aip,jcp,
 amsmath,amssymb,
reprint,%
10pt,%
]{revtex4-2}

\usepackage{graphicx}
\usepackage{dcolumn}
\usepackage{bm}

\usepackage[utf8]{inputenc}
\usepackage[T1]{fontenc}
\usepackage{etoolbox}
\usepackage{mathrsfs}  
\usepackage{amsmath}
\usepackage{amsthm}
\usepackage{amssymb}
\usepackage[urlcolor=white]{hyperref}
\usepackage{mathrsfs}
\usepackage[capitalise]{cleveref}
\usepackage{float}
\usepackage{xcolor}
\usepackage{braket}
\usepackage{comment}

\newcommand{\defgr}{\mathrel{\mathop:\!\!=}}
\newcommand{\norm}[1]{\lVert#1\rVert}
\newcommand{\Norm}[1]{\left\lVert#1\right\rVert}

\DeclareOldFontCommand{\sf}{\normalfont\sffamily}{\mathsf}

\newtheorem{algorithm}{Algorithm}

\makeatletter
\def\@email#1#2{%
 \endgroup
 \patchcmd{\titleblock@produce}
  {\frontmatter@RRAPformat}
  {\frontmatter@RRAPformat{\produce@RRAP{*#1\href{mailto:#2}{#2}}}\frontmatter@RRAPformat}
  {}{}
}%
\makeatother
\begin{document}

\preprint{AIP/123-QED}

\title[Sampling strategies for the Herman--Kluk propagator]{Sampling strategies for the Herman--Kluk propagator of the wavefunction}
\author{Fabian Kr\"oninger}
\email{fabian.kroninger@epfl.ch}
 \affiliation{Laboratory of Theoretical Physical Chemistry, Institut des Sciences et Ingénierie Chimiques, Ecole Polytechnique Fédérale de Lausanne (EPFL), CH-1015 Lausanne, Switzerland}
\author{Caroline Lasser}%
 \email{classer@ma.tum.de}
\affiliation{ 
Zentrum Mathematik, Technische Universit\"at M\"unchen, Germany.
}%

\author{Ji\v{r}\'{\i} J.~L. Van\'{\i}\v{c}ek}
\email{jiri.vanicek@epfl.ch}
\affiliation{Laboratory of Theoretical Physical Chemistry, Institut des Sciences et Ingénierie Chimiques, Ecole Polytechnique Fédérale de Lausanne (EPFL), CH-1015 Lausanne, Switzerland}

\date{\today}

\begin{abstract}
When the semiclassical Herman--Kluk propagator is used for evaluating quantum-mechanical observables or time-correlation functions, the initial conditions for the guiding trajectories are typically sampled from the Husimi density. 
Here, we employ this propagator to evolve the wavefunction itself. 
We investigate two grid-free strategies for the initial sampling of the Herman--Kluk propagator applied to the wavefunction and validate the resulting time-dependent wavefunctions evolved in harmonic and anharmonic potentials. 
In particular, we consider Monte Carlo quadratures based  either on the initial Husimi density or on its square root as possible and most natural sampling densities. 
We prove analytical convergence error estimates and validate them with numerical experiments on the harmonic oscillator and on a series of Morse potentials with increasing anharmonicity. 
In all cases, sampling from the square root of Husimi density leads to faster convergence of the wavefunction.
\end{abstract}
\maketitle
\section{Introduction}\label{Chapter:1}
Semiclassical initial value representation techniques\cite{Miller:1970, Miller:2001} have evolved into useful tools for calculations of the dynamics of atoms and molecules.\cite{book_Heller:2018}
Frozen Gaussians and their superposition were introduced\cite{Heller:1981} by Heller in 1981 as an extension to the thawed Gaussian approximation\cite{Heller:1975} in order to capture nonlinear spreading of the wavepackets. 
Herman and Kluk justified the frozen-Gaussian ansatz and introduced a refined approximation,\cite{Herman_Kluk:1984,Herman:1986} now known as the Herman--Kluk propagator, 
which contains an additional prefactor that rigorously compensates for the fixed width of these frozen Gaussians.

\begin{figure}
    \includegraphics{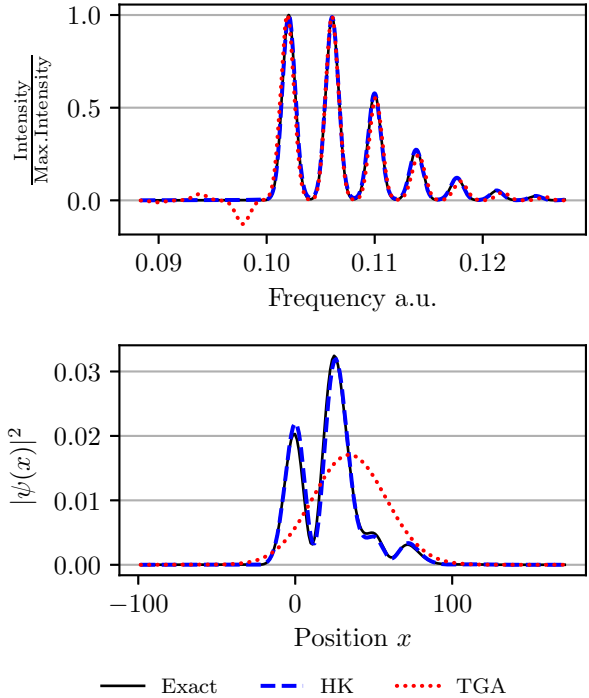}
    \caption{Upper panel: Spectra of a Morse potential evaluated using the exact quantum dynamics, Herman--Kluk (HK) propagator, and thawed Gaussian approximation (TGA). Both approximations yield accurate results.
    Lower panel: Position density at time $t\approx 392$ fs propagated in the same Morse potential. In contrast to the Herman--Kluk propagator, the thawed Gaussian approximation does not capture interference between faster and slower components of the wavepacket. For more details, see the last paragraph of Sec.~IV.}
    \label{fig:Introduction}
\end{figure}

In the spirit of the semiclassical initial value representation, the Herman--Kluk propagator avoids the root search problem.\cite{Miller:2001,Kay:2005} Because of its high accuracy, this propagator belongs among the most successful semiclassical approximations\cite{Kay:1994a,Walton_Manolopoulos:1996,Garashchuk_Tannor:1996,Thoss_Wang:2004,Spanner_Batista:2005,Tatchen_Pollak:2009,Ceotto_Atahan:2009}  and has been derived in many different ways.\cite{Kay:1994b,Miller:2002,Miller:2002a,Shalashilin_Child:2004a,Deshpande_Ezra:2006} There are observables, such as low-resolution vibronic spectra in mildly anharmonic systems, which can be well described by the more efficient thawed Gaussian approximation \cite{Tannor_Heller:1982,Begusic_Vanicek:2018a} and other single-trajectory methods.\cite{Hagedorn:1980,Lee_Heller:1982,Hagedorn:1998,Coalson_Karplus:1990,Begusic_Vanicek:2019,Prlj_Vanicek:2020} In chaotic and other systems, however, increased anharmonicity leads to wavepacket splitting and nontrivial interference effects. In such situations, the single-trajectory methods break down, whereas the Herman--Kluk propagator often remains accurate and, at the least, provides a qualitatively correct insight (see, e.g., \cref{fig:Introduction}). 

The multi-trajectory nature of the Herman--Kluk propagator is not only its advantage, but also its bottleneck. In particular, converged computations might require an extremely large number of trajectories. For this reason, several groups designed various methods, whose goal is reducing the number of trajectories required by the Herman--Kluk propagator. These include, e.g., Filinov filtering,\cite{Filinov:1986,Makri_Miller:1987, Makri_Miller:1988, Walton_Manolopoulos:1996}, time-averaging,\cite{Elran_Kay:1999a,Kaledin_Miller:2003,Buchholz_Ceotto:2016,Buchholz_Ceotto:2018}, semiclassical interaction picture,\cite{Shao_Makri:2000,Petersen_Pollak:2015} multiple-coherent states,\cite{Ceotto_Atahan:2009a}, hybrid dynamics,\cite{Grossmann:2006,Goletz_Grossmann:2009,Grossmann:2016} mixed quantum-semiclassical dynamics,\cite{Antipov_Ananth:2015,Church_Ananth:2017,Malpathak_Ananth:2022} and many others, which achieve the reduction in the number of trajectories by applying one of several possible further approximations.  

Yet, the number of guiding trajectories can already be significantly reduced simply by choosing the sampling density of the initial conditions wisely. Although the acceleration of the convergence may be smaller than with the previously mentioned methods, the advantage of this approach is that the converged results agree exactly with the converged results of the original Herman--Kluk propagator. For this reason, most calculations of observables, time-correlation functions, and wavepacket autocorrelation functions, i.e., quantities quadratic in the wavefunction, correctly employ sampling from the Husimi density of the initial state.\cite{Miller:2001,Pollak_Liu:2022} In contrast, the choice of the optimal sampling density for the Herman--Kluk wavefunction itself has not been analyzed in detail.

The goal of this work is, therefore, to analyze the convergence of the Herman--Kluk wavefunction for different initial sampling strategies and to understand the convergence error as a function of time.
Specifically, we investigate two mesh-free discretization approaches for the initial sampling---first analytically and then  numerically on the examples of harmonic and Morse oscillators with increasing anharmonicity.

The remainder of this paper is organized as follows. 
In the next section we briefly introduce the Herman--Kluk propagator and its components necessary for numerical computations. We define the initial sampling densities and specify the algorithm used for numerical experiments.
In the main Sec.~3, we analyze the errors at the initial and final times due to the phase-space discretization. In particular, we prove that in the harmonic oscillator this error is a periodic function of time. Our numerical experiments in Sec.~4 confirm the theoretical error estimates and provide further insights into the anharmonic evolution generated by Morse potentials, 
which are not accessible to explicit analytical calculations.

\section{Discretising the Herman--Kluk propagator}\label{Chapter:2}
\subsection{Herman--Kluk propagator}
Evolution of a quantum state $\ket{\psi(t)}$ is governed by the time-dependent Schr\"odinger equation
\begin{equation} \label{SchrodingerEquation}
i\hbar \frac{d}{dt}\Ket{\psi(t)}=\hat{H}\Ket{\psi(t)},
\quad \Ket{\psi(0)}=\Ket{\psi_0}
\end{equation}
where $\hbar$ is the reduced Planck constant and $\hat{H}$ is the Hamiltonian operator. Here, we assume the Hamiltonian to be the time-independent operator 
\begin{equation} \label{TI_Hamiltonian}
\hat{H} = \frac{1}{2m} \hat{p}^T \cdot \hat{p} +V(\hat{q}),
\end{equation} 
where $m$ is the mass (after mass-scaling coordinates), and $q$ and $p$ are $D$-dimensional position and momentum vectors. Under the usual regularity and growth assumptions on the potential energy function $V$, the spectral-theorem \cite{Hall:2013} provides for all times $t\in\mathbb{R}$ a well-defined unitary propagator
\begin{equation}
\hat{U}_t\defgr \exp{({-i\hat{H} t/\hbar})},
\end{equation} in terms of which the solution of \eqref{SchrodingerEquation} can be expressed as
\begin{equation}
\Ket{\psi(t)}=\hat{U}_t\Ket{ \psi_0}
\end{equation} 
for all square integrable initial data $\psi_0(x)=\langle x \ket{\psi_0}\in L^2(\mathbb{R}^D)$. 

Solving the Schr\"odinger equation numerically is notoriously difficult for various reasons. With respect to the atomic scale, the nuclear mass is rather large, so that the presence of the factor $m^{-1}$ in the kinetic energy operator induces highly oscillatory motion both with respect to time and space. More importantly, for most molecular systems the dimension $D$ of the configuration space is so large that grid-based integration methods are very expensive if not infeasible. Therefore, one often resorts to mesh-free discretization methods or semiclassical approximations, both of which alleviate this ``curse of dimensionality'' at least partially. 

The semiclassical Herman--Kluk propagator utilizes frozen Gaussian functions
\begin{align}
\begin{split}
&g_z^\gamma(x)  = \left(\frac{\det{\gamma}}{\pi^D \hbar^D}\right)^{1/4}
\\ & \times \exp{\left\{ -\left[(x-q)^T \cdot \gamma \cdot  (x-q)/2+ip^T \cdot(x-q)\right]/\hbar\right\}}
\end{split}
\end{align}
 with a centre at the phase-space point $z=(q,p)\in\mathbb{R}^{2D}$ and with a fixed, 
real, symmetric, positive-definite width-matrix $\gamma\in\mathbb{R}^{D\times D}$. The frozen Gaussians $\{\ket{g_z^\gamma}: z\in\mathbb{R}^{2D}\}$ form an over-complete subset of the Hilbert space of square integrable functions. They have the striking property that any $\ket{\psi}\in L^2(\mathbb{R}^D)$ can be decomposed as\cite{martinez}
\begin{equation}
\Ket{\psi}= \int_{\mathbb{R}^{2D}}\langle g_z^\gamma \vert \psi\rangle \Ket{g_z^\gamma} d\nu,
\end{equation} 
with the scaled phase-space measure $d \nu = dz/(2 \pi \hbar)^D.$
Using this decomposition for our solution of the time-dependent Schr\"odinger equation, we obtain
\begin{equation} \label{Decomposition}
\hat{U}_t\Ket{\psi_0}=\int_{\mathbb{R}^{2D}}  \langle g_z^\gamma \vert \psi_0 \rangle \hat{U}_t\Ket{ g_z^\gamma} \,d\nu.
\end{equation} 
Based on the continuous superposition \eqref{Decomposition}, the state obtained by applying the Herman--Kluk propagator to $\ket{\psi_0}$ can be expressed as
\begin{equation} 
\label{Herman-Kluk}
\Ket{\hat{U}^{\rm HK}_t \psi_0} \defgr \int_{\mathbb{R}^{2D}} R(t,z) e^{\frac{i}{\hbar} S(t,z)} \langle g_z^\gamma\vert \psi_0 \rangle \ket{g_{z(t)}^\gamma} d\nu,
\end{equation} 
where $z(t)=(q(t), p(t))$ is the solution to the underlying classical Hamiltonian system
\begin{equation} \label{CHS}
\dot w =J \cdot \nabla h(w),\hspace{0.5cm} w(0)=z,
\end{equation} 
for the Hamiltonian function $h(q,p)=T(p)+V(q)$, where \begin{equation}
    J=\begin{pmatrix}
0 & \mathrm{Id}_D \\ -\mathrm{Id}_D & 0
\end{pmatrix}\in\mathbb{R}^{2D\times 2D}
\end{equation} is the standard symplectic matrix. $S$ denotes the classical action integral 
\begin{equation}
S(t,z )= \int_0^t \left[\frac{d}{d\tau} q(\tau)\cdot p(\tau) -h(z(\tau))\right]d\tau 
\end{equation} which solves the initial value problem
\begin{align} \label{ClassicalAction} 
\dot{S}(t,z)=T(p(t))-V(q(t)), \hspace{0.5cm} S(0,z)=0,
\end{align} for all $z=(q,p)\in\mathbb{R}^{2D}.$
In \eqref{Herman-Kluk}, the Herman--Kluk prefactor
\begin{align}
\begin{split}
&R(t,z) = 2^{-D/2}
\\&\cdot \det{\left(M_{qq}+ \gamma^{-1} \cdot M_{pp}\cdot \gamma - i M_{qp}\cdot \gamma +i \gamma^{-1}\cdot  M_{pq}\right)^{1/2}}
\end{split}
\end{align} depends on the matrices 
\begin{equation}
    \label{eq:components_of_M}
    M_{\alpha \beta} = \partial \alpha_t/\partial\beta_0\in\mathbb{R}^{D\times D}
    \ \ \ (\alpha,\beta \in \{q,p\}),    
\end{equation}
i.e., the four $D \times D$ block components of the $2D \times 2D$ stability matrix $M.$ 
Stability matrix, defined as the Jacobian $M(t)=\partial z(t)/\partial z$ of the flow map, is the solution to the variational equation 
\begin{equation} \label{VariationalEquation} 
\dot{M}(t)=J\cdot \mathrm{Hess} \, h(z(t)) \cdot M(t), \hspace{0.5cm} M(0)=\textrm{Id}_{2D},
\end{equation} with $\mathrm{Hess} \, h$ the Hessian of the Hamiltonian function $h$. 

The Herman--Kluk approximation has been mathematically justified in different works.\cite{Kay:2006,Lasser_Lubich:2020,Swart_Rousse:2008, Robert2010} It has been shown that
the exactly evolved quantum state is approximated by the state 
\begin{align}
    \ket{\psi^{\rm HK}(t)} := \ket{\hat{U}^{\rm HK}_t\psi_0},
\end{align}
obtained by applying the Herman--Kluk propagator to the initial state $\ket{\psi_0}$,
with an error of the order of $\hbar.$ More precisely,
\begin{equation} \label{Theorem1}
\sup_{t\in[0,T]}\Norm{ \hat{U}_t\Ket{\psi_0} -\ket{\psi^{\rm HK}(t)}}_{L^2} \leq C(T)\hspace{0.1cm} \hbar,
\end{equation} for all initial data $\ket{\psi_0}$ of norm one, where $T>0$ is a fixed time and $C(T)>0$ is a constant independent of $\hbar.$ If the potential is at most harmonic, then the approximation is exact. \cite{Lasser_Lubich:2020}

\subsection{Discretisation} \label{K1.2}
Evolving the wavefunction~\eqref{Herman-Kluk} with the Herman--Kluk propagator requires evaluating an integral over the phase space $\mathbb{R}^{2D}$ and the overlap of the initial state with a frozen Gaussian. Furthermore, the algorithm needs to propagate the trajectories $z(t)=(q(t),p(t)),$ the classical action $S(t,z)$, and the Herman--Kluk prefactor $R(t,z)$ according to Hamilton's equations of motion for all chosen quadrature points $z\in\mathbb{R}^{2D}.$ The latter can be achieved using  symplectic integration methods to  preserve also the symplectic structure of the classical Hamiltonian system. Due to the curse of dimensionality, for high $D$ the integral on $\mathbb{R}^{2D}$ must be evaluated using mesh-free discretization, such as Monte Carlo methods. \cite{Caflisch:1998}

To evaluate the Herman--Kluk wavefunction by Monte Carlo sampling, we rewrite the integral from \cref{Herman-Kluk} as
\begin{equation}
    \ket{\psi^{\rm HK}(t)} = \left\langle r(z) \Ket{\phi(t,z)}\right\rangle_{\rho(z)},
\end{equation}
where we introduced the notation
\begin{align}
    \begin{split}
    \left\langle  \Ket{\psi(t,z)}\ \right\rangle_{\rho(z)} &\defgr \int_{\mathbb{R}^{2D}} \Ket{\psi(t, z)} \,d\mu
    \\&= \int_{\mathbb{R}^{2D}} \Ket{\psi(t, z)} \rho(z)\,d\nu
    \end{split}
\end{align}
for the phase-space average of $\ket{\psi(t,z)}$ with respect to a probability measure $\mu$ with density $\rho(z) = d\mu/d\nu$ and defined a time-dependent state 
\begin{equation}
   \Ket{\phi(t,z)} :=  R(t,z)e^{\frac{i}{\hbar}S(t,z)}\ket{g^\gamma_{z(t)}},
   \label{eq:phi_t_z}
\end{equation}
i.e., a propagated frozen Gaussian multiplied by the Herman--Kluk and phase prefactors, and a time-independent function
\begin{equation}
    r(z) \defgr \langle g_z^\gamma \vert \psi_0\rangle \,\rho(z)^{-1}.
\end{equation}
The Monte Carlo estimator is then given by
\begin{align} \label{MonteCarloEstimator}
    \Ket{\psi_N(t)}=\frac{1}{N}\sum_{j=1}^N \Ket{\psi(t,z_j)},
\end{align}
where $\psi(t,z_j)$ is the state obtained from initial condition $z_j$ and $z_1, z_2 , \dots , z_N$ are sampled from probability density $\rho(z)$. 

As for any importance sampling, there are infinitely many ways to decompose the time-independent part of the phase space integrand in \cref{Herman-Kluk} into the product $\langle g_z^\gamma \vert \psi_0\rangle = r(z) \rho(z)$
of a prefactor $r$ with a normalized sampling density $\rho$.
If one computes observables and correlation functions, which are quadratic in the initial state, $\rho(z)$ is typically taken to be the Husimi probability density
\begin{align}
    \rho_{\rm H}(z)\defgr |\langle g_z^\gamma \vert \psi_0\rangle|^2 \label{rho_Husimi}
\end{align}
of the initial state $\ket{\psi_0}$.  For Husimi distribution, the prefactor $r(z)$ becomes
\begin{equation}
    \label{eq:r_Husimi}
    r_{\rm H}(z) = \langle \psi_0 \vert g_z^\gamma\rangle^{-1}.
\end{equation}
However, since the wavefunction evolved with the Herman--Kluk propagator is first- and not second-order in $\ket{\psi_0}$, it is natural to also test the square root of the Husimi density and to consider
the probability density
\begin{equation} 
    \label{Sqrt-Husimi}
    \tilde{\rho}(z) \defgr \frac{|\langle g_z^\gamma \vert \psi_0\rangle|}{\int_{\mathbb{R}^{2D}}|\langle g_w^\gamma \vert \psi_0\rangle| d\nu} 
\end{equation} 
with a bounded prefactor
\begin{equation}
    \tilde{r}(z) \defgr \langle g_z^\gamma \vert \psi_0\rangle \,\tilde{\rho}(z)^{-1}.
\end{equation}

The two probability densities $\rho_{\rm H}(z)$ and $\tilde{\rho}(z)$ can now be used to compute the phase-space integral by Monte Carlo integration. To sum up, we have two cases, in which we evaluate $\ket{\psi^{\rm HK}(t)}$ either as
\begin{align} \label{Husimi}
\ket{\psi^{\rm HK}(t)} &= \left\langle r_{\rm H}(z) \Ket{\phi(t,z)}\right\rangle_{\rho_{\rm H}(z)} \tag{Case H}
\end{align}
or as
\begin{align} \label{sqrt-Husimi}
\ket{\psi^{\rm HK}(t)} 
=\left\langle \tilde{r}(z) \Ket{\phi(t,z)}\right\rangle_{\tilde{\rho}(z)}. \tag{Case sqrt-H}
\end{align}
In general, the integral over $\mathbb{R}^D$ defining the overlap of the initial wavefunction with a Gaussian has to be computed by numerical quadrature. However, for important specific cases analytical formulas are available.   
If the initial wavefunction is a Gaussian wavepacket $\ket{\psi_0}=\ket{g_{z_0}^\gamma}$ centred at some phase-space point $z_0=(q_0,p_0)\in\mathbb{R}^{2D}$, then
\begin{align} 
    \label{GaussDecomposition}
    \begin{split}
    \langle g_z^\gamma \vert \psi_0\rangle_{L^2(\mathbb{R}^{D})} 
     &= \exp{\left[-\frac{1}{4\hbar} (z-z_0)^T\cdot \Sigma_0\cdot(z-z_0)\right]}  \\ 
     &\cdot \exp{\left[\frac{i}{2\hbar}(p+p_0)^T\cdot(q-q_0)\right]},
     \end{split}
\end{align}
where $\Sigma_0\defgr \begin{pmatrix}
\gamma & 0 \\ 0 & \gamma^{-1}
\end{pmatrix}$ is the matrix containing the width parameters of the initial coherent state. Then, the Husimi density is given by
\begin{equation}
    \label{eq:Husimi_GWP}
    \rho_{\rm H}(z) = \exp{\left[-\frac{1}{2\hbar} (z-z_0)^T\cdot\Sigma_0\cdot(z-z_0)\right]},
\end{equation}
whereas the second approach provides the density
\begin{equation}
    \label{eq:Sqrt-Husimi_GWP}
    \tilde{\rho}(z)= 2^{-D}\exp{\left[-\frac{1}{4\hbar} (z-z_0)^T\cdot\Sigma_0\cdot(z-z_0)\right]}    
\end{equation} 
and prefactor
\begin{equation}
    \label{eq:r_Sqrt-Husimi_GWP}  
    \tilde{r}(z)=2^D\exp{\left[\frac{i}{2\hbar}(p+p_0)^T\cdot(q-q_0)\right]}.
\end{equation}

\subsection{Summary of the numerical algorithm} \label{K1.3}
Taking into account all the previous considerations, we slightly extend the natural numerical algorithm (described, e.g., in Ref.~\onlinecite[Sec. 4]{Lasser_Sattlegger:2017}]) for finding the Herman--Kluk approximation to the wavefunction at time $t$: 
\begin{algorithm}{\bf{(Herman--Kluk propagation)}}  \label{Alg1} 
\begin{enumerate}
\item Draw independent samples $z_1, \dots , z_N\in \mathbb{R}^{2D}$ from a distribution with density $\rho_{\rm H}(z)$ or $\tilde{\rho}(z)$ given by Eqs.~(\ref{eq:Husimi_GWP}) and (\ref{eq:Sqrt-Husimi_GWP}).
\item For all $j\in \{0,1, \dots , N \}:$
	\begin{enumerate}
	\item[2.1] Set initial values $z(0)=z_j,$ $M(0)=\textrm{Id}_{2D}$ and $S(0)=0.$
	\item[2.2] Compute approximate solutions to \cref{CHS,ClassicalAction,VariationalEquation} up to time $t$ with a symplectic integration method\cite{Brewer_Hulme:1997} based on the St\"ormer--Verlet scheme. \cite{book_Hairer_Wanner:2006,hairer_lubich_wanner:2003}
	\item[2.3] Compute the Herman--Kluk prefactor $R(t,z_j)$  from $M(t)$ while choosing the correct branch of the complex square root,\cite{Stewart_Tall:2018} which guarantees continuity of $R(t,z_j)$ as a function of $t$.
	\end{enumerate}
\item Calculate $\ket{\psi_N(t)}$ by means of formula \eqref{MonteCarloEstimator} with $\ket{\psi(t,z_j)}$ replaced with either
 \begin{align}
    &\Ket{\psi_{\rm H}(t,z_j)}=r_{\rm H}(z_j) R(t,z_j)e^{\frac{i}{\hbar}S(t,z_j)} \Ket{g^\gamma_{z_j(t)}}
    \\ \text{or}& \notag
    \\&\Ket{\tilde{\psi}(t,z_j)}=\tilde{r}(z_j)R(t,z_j)e^{\frac{i}{\hbar}S(t,z_j)}\Ket{g^\gamma_{z_j(t)}},
\end{align}
   where $r_{\rm H}(z_j)$ and $\tilde{r}(z_j)$ are given by Eqs.~(\ref{eq:r_Husimi}), (\ref{GaussDecomposition}), and (\ref{eq:r_Sqrt-Husimi_GWP}).
\end{enumerate} $\hfill \qedsymbol$
\end{algorithm}
We note that \cref{CHS,ClassicalAction,VariationalEquation} can be evaluated simultaneously with a single numerical integrator. To increase the accuracy of the time integration one can use higher-order composition methods.\cite{Kahan_Li:1997,Suzuki:1992} They increase the order of the time integrator, but also its numerical cost. A higher-order time integrator is not necessarily useful since the phase-space error occurring from the Monte Carlo quadrature usually dominates the time integration error.

\section{Phase space error analysis} \label{Chapter:3}
Algorithm~\ref{Alg1} relies on the discretisation of the phase-space integrals and of the system of ordinary differential equations. Here we focus only on the phase-space discretisation error.

\subsection{Moments of the integrand}
To assess the accuracy of the Monte Carlo estimator \eqref{MonteCarloEstimator}, we examine the moments of the two different integrands: $\ket{\psi_{\rm H}(t)}$ and $\ket{\tilde{\psi}(t)}.$
First, we observe that for any $s>0$,
\begin{align}
\mathbb{E}\big[\norm{\tilde{\psi}(t)}^s\big] = 
\int_{\mathbb{R}^{2D}} |R(t,z)|^s \, |\tilde{r}(z)|^s\, d\tilde{\mu}(z) <\infty,
\end{align}
since $\tilde{r}(z)$ and the Herman--Kluk prefactor $R(t,z)$ are both bounded functions. For a discussion of the boundedness of $R(t,z)$, see \Cref{Appendix:1}. 
In contrast, 
\begin{align}
\begin{split}
\mathbb{E}\left[\norm{\psi_{\rm H}(t)}^s\right] &= 
 \int_{\mathbb{R}^{2D}} |R(t,z)|^s\, |r_{\rm H}(z)|^s  
\ d\mu_{\rm H}(z)\\
&= \int_{\mathbb{R}^{2D}} |R(t,z)|^s\  
|\langle g_z^\gamma \vert \psi_0\rangle|^{2-s}\ d\nu
\end{split}
\end{align}
ceases to be finite for $s \ge 2$. However, both integrands have a finite first moment, so that the Strong Law of Large Numbers (Theorem 2.4.1 in Ref.~\onlinecite{durrett:2019}) provides convergence of the estimator,
\begin{align}
\Ket{\psi_N(t)}\to\ket{\psi^{\rm HK}(t)}\quad 
\text{as}\quad   N\to \infty,
\end{align}
with probability one. The non-existing second moment for \eqref{Husimi} results in a slightly worse convergence rate  than the one for \eqref{sqrt-Husimi}. Numerical results in Sec. \ref{Chapter:4} confirm this expectation.

\subsection{Mean squared error}
For \eqref{sqrt-Husimi} the second moment is finite, so that 
the mean squared error of the Monte Carlo estimator (\ref{MonteCarloEstimator}) is well-defined and satisfies
\begin{equation} \label{Prop3.1}
\mathbb{E}\big[\norm{\psi_{N}(t)-\psi^{\rm HK}(t)}^2\big]=\frac{\mathbb{V}\big[\ket{\tilde{\psi}(t)}\big]}{N},
\end{equation} 
where the expectation value and the variance are with respect to the density $\tilde{\rho}(z)$. Moreover,  \cite{Lasser_Sattlegger:2017}
\begin{align} \label{VarianceCaseA}
\begin{split}
&\mathbb{V}\big[\ket{\tilde{\psi}(t)}\big] 
\\&= \int_{\mathbb{R}^{2D}} \norm{\tilde{\psi}(t)-\mathbb{E}[\tilde{\psi}(t)]}^2  \, d\tilde{\mu}(z) 
\\&= \int_{\mathbb{R}^{2D}} |R(t,z)|^2 \,|\tilde{r}(z)|^2\, d\tilde{\mu}(z)-\norm{\psi^{\rm HK}(t)}^2.
\end{split}
\end{align} 
In the special case of an initial Gaussian initial wavepacket $\ket{\psi_0} = \ket{g_{z_0}^\gamma}$, this simplifies to 
\begin{align}
\mathbb{V}\big[\ket{\tilde{\psi}(t)}\big]= 4^D\int_{\mathbb{R}^{2D}} |R(t,z)|^2\, d\tilde{\mu}(z)-\norm{\psi^{\rm HK}(t)}^2,
\end{align}
and at initial time $t=0$ we obtain
\begin{align} \label{VarianceCaseAT0}
&\mathbb{V}\big[\ket{\tilde{\psi}(0)}\big] =4^D-1,
\end{align}
since the Herman--Kluk prefactor satisfies $R(0,z)=1$. 

For a more general assessment of the variance,  numerical experiments in Sec.~\ref{Chapter:4} consider the error between the approximations with $N$ and $2N$ samples. By the linearity of the expectation value and the triangle inequality, this error can be estimated by
\begin{align}
\begin{split}
    &\mathbb{E}\big[\norm{\psi_N(t)-\psi_{2N}(t)}^2\big] \\
    &\leq \mathbb{E}\big[\norm{\psi_N(t)- \psi^{\rm HK}(t)}^2\big] 
    +\mathbb{E}\big[\norm{\psi^{\rm HK}(t)-\psi_{2N}(t)}^2\big] 
    \\&= \frac{3}{2} \frac{\mathbb{V}\big[\Ket{\psi(t)}\big]}{N}.
\end{split}
\end{align}

Note that \cref{Prop3.1} gives only the expected convergence error, i.e., the convergence error averaged over infinitely many independent simulations, each using $N$ trajectories. The actual convergence error for any specific simulation with $N$ trajectories may deviate from this analytical estimate substantially due to statistical noise. Nevertheless, in \Cref{Appendix:2}, we explain how \cref{Prop3.1} can also provide a rigorous lower bound and asymptotic estimate of the number of trajectories needed for convergence of a single simulation.
\subsection{Other sampling densities}
For the special case of an initial Gaussian wavepacket $\ket{\psi_0} = \ket{g_{z_0}^\gamma}$, the 
two proposed sampling densities $\rho_{\rm H}(z)$ and $\tilde{\rho}(z)$ belong to a family of normal distributions with probability density functions
\begin{align}
\rho_{a}(z) = (a\pi\hbar)^{-D}\exp{\left[-\frac{1}{a\hbar} (z-z_0)^T\cdot\Sigma_0\cdot(z-z_0)\right]}
\end{align}
with $a\ge 2$. 
In the spirit of importance sampling, the Herman--Kluk wavefunction can accordingly be written as a phase-space average
\begin{align}
\ket{\psi^{\rm HK}(t)} =\left\langle \ket{\psi_a(t,z)} \right\rangle_{\rho_{a}(z)}.
\end{align}
At time $t=0$, the norm of the integrand satisfies
\begin{align}
\begin{split}
&\norm{\psi_a(0,z)} \\
&=\left(\frac{a}{2}\right)^{D}
\exp{\left[-\left(\frac{a-4}{4a\hbar}\right) (z-z_0)^T\cdot\Sigma_0\cdot(z-z_0)\right]},
\end{split}
\end{align}
which implies for the variance
\begin{align}
\begin{split}
&\mathbb{V}[\Ket{\psi_a(0)}] \\
& = \left(\frac{a}{4\pi\hbar}\right)^{D} 
\int_{\mathbb{R}^{2d}} \exp{\left[-\left(\frac{a-2}{2a\hbar}\right) |z|^2\right]} dz - 1 \\ &=
\left(\frac{a^2}{2(a-2)}\right)^{D}-1.
\end{split}
\end{align}
For $a>2$, the function $a\mapsto a^2/(2a-4)$ attains its minimum at $a=4$, which corresponds to the sampling density $\tilde{\rho}$. In other words, \eqref{sqrt-Husimi} is optimal as far as the mean squared error is concerned.

\subsection{Harmonic motion}\label{Subsection:3.2}
The one-dimensional harmonic oscillator 
\begin{align}
\hat H = -\frac{\hbar^2}{2m}\frac{d^2}{dx^2} + \frac{m\omega^2}{2} x^2
\end{align}
is one of the rare examples for which explicit expressions for the solution of the Schr\"odinger equation and for the Herman--Kluk prefactor exist. For an initial Gaussian wavepacket $\ket{\psi_0} = \ket{g^\gamma_{z_0}}$ with position and momentum $q_0,p_0\in\mathbb{R},$ the exact wavefunction is given by\cite{book_Tannor:2007}
\begin{align} \label{Harmonic_exact}
    \psi_{\rm ex}(x,t)= \exp{\left\{\frac{i}{\hbar} \left[ \frac{\alpha_t}{2}(x-q_t)^2+p_t(x-q_t)+ \beta_t \right] \right\} }
\end{align}
where 
\begin{align}
    \alpha_t &=b \frac{\alpha_0\cos(\omega t)-b \sin(\omega t)}{b \cos (\omega t) + \alpha_0\sin(\omega t)}, \label{Harmonic_exact_width}
    \\q_t&= q_0 \cos(\omega t) + \frac{p_0}{b}\sin(\omega t), \label{Harmonic_exact_position}
    \\p_t&=p_0 \cos(\omega t) - b q_0 \sin(\omega t) \text{ and } \label{Harmonic_exact_momentum}
    \\ \beta_t &= \beta_0+\frac{1}{2}\left[q_tp_t-q_0p_0 + i\hbar \ln{\left(\frac{z_t}{b}\right)}\right] \label{Harmonic_exact_gamma}
\end{align}
with the abbreviations $z_t=b\cos(\omega t)+ \alpha_0 \sin(\omega t),$ $\alpha_0=i\gamma$ and $b=m\omega.$ Here, $\beta_0$ includes the normalization constant for the wavefunction at time $t=0.$ The classical action is given by
\begin{align} \label{Harmonic_exact_action}
\begin{split}
    &S(t,q_0,p_0)
    \\&=\frac{1}{2}\sin(\omega t)\left[ \left(\frac{p_0^2}{b}-bq_0^2 \right) \cos(\omega t) - q_0p_0 \sin(\omega t) \right]. 
    \end{split}
\end{align}
The four components of stability matrix $M$ can be obtained, from their definition (\ref{eq:components_of_M}), by differentiating expressions for $q_t$ and $p_t$ with respect to $q_0$ and $p_0$, namely
\begin{align}
M(t)= \begin{pmatrix}
\cos{\left(\omega t\right)} & b^{-1}\sin{\left(\omega t\right)} \\ - b \sin{\left(\omega t \right)} & \cos{\left(\omega t \right)} 
\end{pmatrix}.
\end{align}
The Herman--Kluk prefactor satisfies
\begin{align}
    R(t) = \left\{\frac{1}{2} \left[ 2\cos{\omega t}-i\sin(\omega t) \left(\frac{\gamma}{b}+ \frac{b}{\gamma} \right)\right] \right\}^{1/2}, \label{Harmonic_exact_hkprefactor}
\end{align}
so that the variance \eqref{VarianceCaseA} can be written as
\begin{align}\label{VarianceHarmonicOverTime}
\begin{split}
    &\mathbb{V}\big[\ket{\tilde{\psi}(t)}\big]
    \\&=2\left[4\cos^2(\omega t) + \left( \frac{\gamma}{b}+ \frac{b}{\gamma} \right)^2 \sin^2(\omega t)\right]^{1/2} -1.
    \end{split}
\end{align}
This implies that for \eqref{sqrt-Husimi} applied to a harmonic oscillator, the mean squared error of our Monte Carlo estimator oscillates with a period of $\pi/\omega$ between
\begin{align}
    3\leq \mathbb{V}\big[\ket{\tilde{\psi}(t)}\big] \leq 2\left(\frac{\gamma}{m\omega}+\frac{m\omega}{\gamma} \right) -1.
\end{align}

\section{Numerical examples} \label{Chapter:4}
In this section, we complement our previous theoretical results with numerical examples. We start by examining how the performance of \Cref{Alg1} depends on the method to  discretise the phase space. 
We explore the time dependence of the variance of the Monte Carlo estimator in one dimension in a harmonic oscillator as well as in a series of increasingly anharmonic Morse potentials. 
The Monte Carlo integration is tested by averaging over $N$ independent, identically distributed samples of initial conditions. We approximate its convergence rate by assuming a power law 
\begin{align} \label{Linear_Fitting}
    F(N) = c N^{-s}
\end{align} dependence of the mean statistical error on the number of samples $N$. The prefactor $c$ and order $s$ of convergence are determined by the linear fit (in the least-squares sense) of the logarithm of Eq.~(\ref{Linear_Fitting}) to the dependence of the logarithm of the statistical error on the logarithm of $N$.

Throughout our numerical examples, we work in atomic units $(\hbar=1),$ mass-scaled coordinates and with an initial state that is a Gaussian wavepacket with phase-space centre $z_0\in\mathbb{R}^{2D}.$

\subsection{Initial phase-space error}

\begin{figure}
    \includegraphics{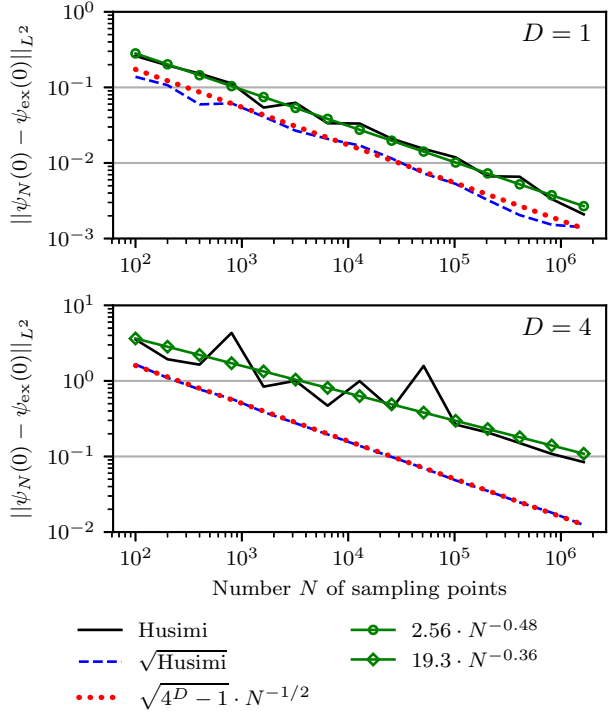}
    \caption{Sampling error of the initial wavefunction in one (upper panel) and four (lower panel) dimensions as a function of the number $N$ of Monte Carlo points. The plot displays the error for sampling from the Husimi density \eqref{Husimi} and its approximated convergence (marked lines) as well as the error for sampling from the square root of the Husimi density \eqref{sqrt-Husimi} and its analytical error estimation (dotted line). }
    \label{fig:Initial_error}
\end{figure}
\begin{figure}
    \includegraphics{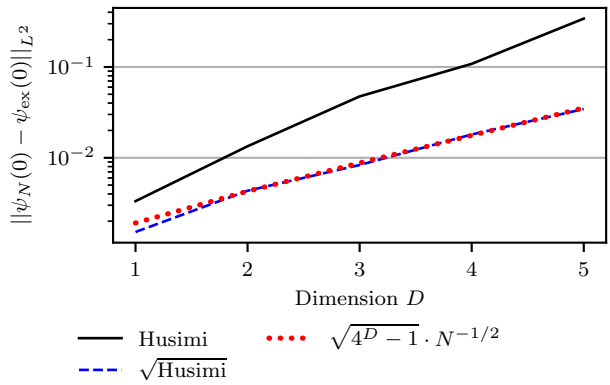}
    \caption{Dependence of the sampling error of the initial wavefunction on dimension $D$ for $N=100\cdot 2^{13} \approx 8\cdot 10^5$ points sampled from either the Husimi density \eqref{Husimi} or its square root  \eqref{sqrt-Husimi}. The analytical error estimate for the latter sampling is shown by the dotted line.}
    \label{fig:Initial_error_over_D}
\end{figure}

We start by considering a spherical initial Gaussian wavepacket $\ket{\psi_{\rm ex}(0)} = \ket{g^\gamma_{z_0}}$ with a width parameter $\gamma=2 \mathrm{Id}_{D}$ in one and four dimensions.
For $D=1$, it is centreed  at $z_0=(-1,0)$ and for $D=4$, at $z_0=-(1,1,1,1,0,0,0,0)$.  \Cref{fig:Initial_error} shows the $L^2$-distance 
\begin{align}
\norm{\psi_N(0) - \psi_{\rm ex}(0)}_{L^2}
\end{align}
of the Monte Carlo estimators
for \eqref{Husimi} and \eqref{sqrt-Husimi} from the exact wavefunction at initial time as a function of the number of Monte Carlo quadrature points. 
We can immediately see that the analytical prediction \eqref{VarianceCaseAT0} of the mean squared error \eqref{Prop3.1} of \eqref{sqrt-Husimi} is fulfilled.
For \eqref{Husimi}, the curve-fitting approximation \eqref{Linear_Fitting} provided $(c,s)=(2.56,-0.48)$ for $D=1$ and $(c,s)=(19.3,-0.36)$ for $D=4.$ It shows that both cases converge to the correct result and that \eqref{sqrt-Husimi} performs slightly better than \eqref{Husimi}.
Additionally, \cref{fig:Initial_error_over_D} displays the $L^2$-distance of the estimators for both cases from the exact wavefunction at initial time as a function of the dimension $D.$ Each wavefunction was approximated with $N\approx 8\cdot 10^5$ trajectories. The analytical prediction \eqref{VarianceCaseAT0} for \eqref{sqrt-Husimi} is realized. Moreover, the error for \eqref{Husimi} increases faster with $D$.

\subsection{Harmonic potential}
\begin{figure}
    \includegraphics{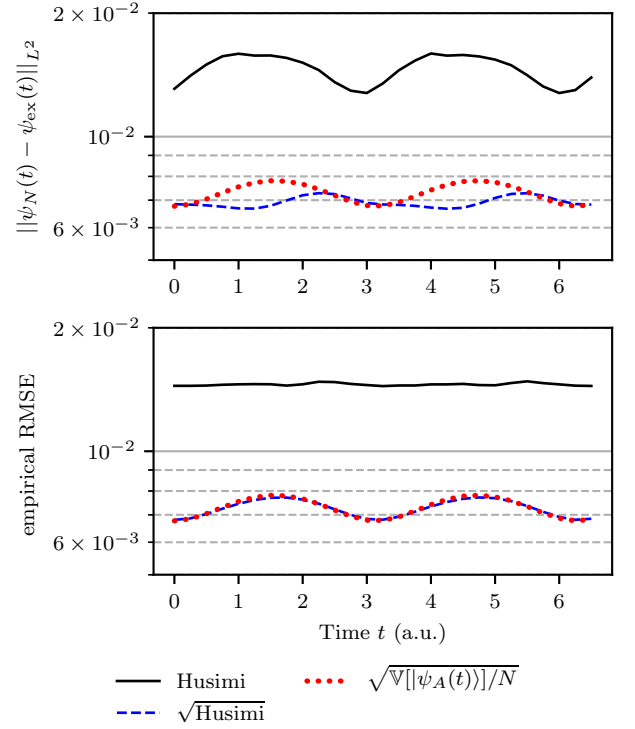}
    \caption{Time dependence of the sampling error of the Herman--Kluk wavefunction propagated in a harmonic oscillator. The upper panel is produced by one independent run with $N=2^{16}=65536$ trajectories, whereas the lower panel is produced by $K=100$ independent runs, each with $N=2^{16}=65536$ trajectories, and averaging the square of the error over the $K$ runs. The analytical error estimate for the sampling from the square root of the Husimi density \eqref{sqrt-Husimi} is shown with the dotted line.
    }
    \label{fig:Harmonic_error}
\end{figure}
To analyze the effect of dynamics on the convergence, let us consider a harmonic potential $V(x)=x^2/2$ in one dimension and explore the dynamics for one full oscillation period. The initial Gaussian wavepacket is still localized in $z_0=(-1,0)$ with width $\gamma=2.$ In Fig.~\ref{fig:Harmonic_error}, we compare the exact wavefunction \eqref{Harmonic_exact} with the numerical realization of the Herman--Kluk propagator which uses the exact solution to $q(t),p(t),S(t,z)$ and $R(t,z)$ stated in  \cref{Harmonic_exact_position,Harmonic_exact_momentum,Harmonic_exact_action,Harmonic_exact_hkprefactor}. Since the Herman--Kluk propagator is exact for harmonic motion and since we supply exact classical trajectories, the observed numerical error is only due to the Monte Carlo integration. 
The upper panel of \cref{fig:Harmonic_error} shows the time dependence of the $L^2$-error 
\begin{align}
\norm{\psi_N(t) - \psi_{\rm ex}(t)}_{L^2},
\end{align}
where the semiclassical wavefunctions were generated using $N=2^{16}$ trajectories. Sampling from the square root of the Husimi density \eqref{sqrt-Husimi} results in approximately twice smaller error than sampling from the Husimi density itself. For \eqref{sqrt-Husimi}, the figure also displays the analytical error estimate derived in \eqref{VarianceHarmonicOverTime}, which  
matches the numerical error up to small statistical noise. 
To remove this statistical noise and to match the analytical estimate \eqref{VarianceHarmonicOverTime} more accurately, in the lower panel of Fig.~\ref{fig:Harmonic_error} we plot the empirical root mean square error (RMSE) \cite{Caflisch:1998} $S_{100}$, where
\begin{align}
S_K \defgr \frac{1}{K} \sum_{j=1}^K \norm{\psi_N^{(j)}(t) - \psi_{\rm ex}(t)}^2, 
\end{align}
$K$ is an integer number of independent simulations (indexed by $j$) and each $\psi_N^{(j)}$ was itself approximated using $N$ independent samples. One can see that
\begin{align}
 \mathbb{E}[S_K]=&\mathbb{E}\left[\norm{\psi_N(t) - \psi_{\rm ex}(t)}^2\right]= \frac{\mathbb{V}\big[\ket{\tilde{\psi}(t)}\big]}{N}.
\end{align}
For $K=1$, one obtains the result represented in the upper panel of Fig.~\ref{fig:Harmonic_error}. For $K \rightarrow \infty,$ due to the Strong law of large numbers, $S_K$ converges almost surely to its expectation value and hence to the analytical error estimate. 

The lower panel of \cref{fig:Harmonic_error} shows the empirical RMSE computed with $K=100$ and $N=2^{16}.$ For \eqref{sqrt-Husimi}, it coincides almost perfectly with the analytical prediction. We note that the error reaches its maximum whenever the wavefunction passes through the bottom of the potential and its minimum when the wavefunction arrives at the turning points.
Even though this analysis makes only sense for finite variance, the figure also displays the empirical RMSE for \eqref{Husimi}, which is nearly constant.
Both panels show a qualitatively similar error evolution for \eqref{Husimi}, however, at a magnitude that is considerably larger than the error for \eqref{sqrt-Husimi}.

\subsection{Morse potential}
\begin{figure}
    \centering
    \includegraphics{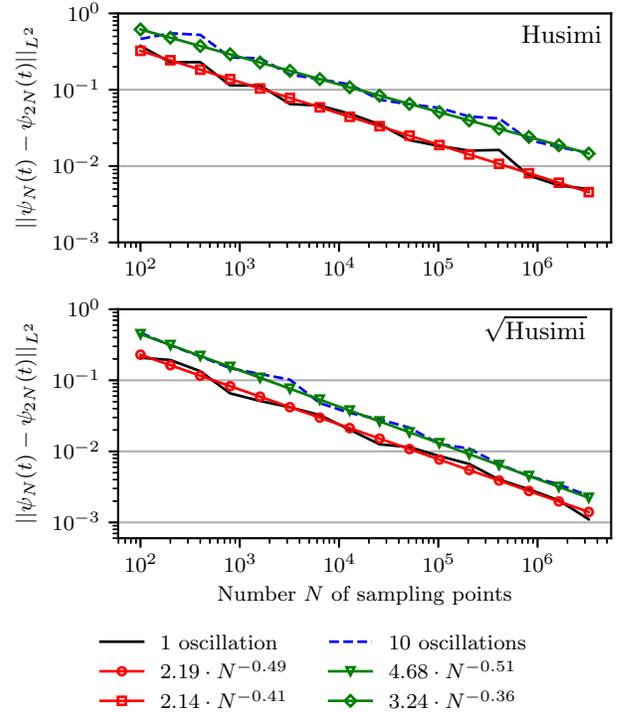}
    \caption{Sampling error between the Herman--Kluk wavefunctions obtained with $N$ and $2N$ Monte Carlo quadrature points as a function of $N$. The wavefunctions are calculated in a Morse potential with anharmonicity parameter $\chi=0.005$ after approximately one (solid line) or ten oscillations (dashed line). The upper panel shows both errors and their approximated convergence rates for \eqref{Husimi}. Similarly, \eqref{sqrt-Husimi} is displayed in the lower panel.}
    \label{fig:Morse_chi_0005}
\end{figure}
\begin{figure}
    \centering
    \includegraphics{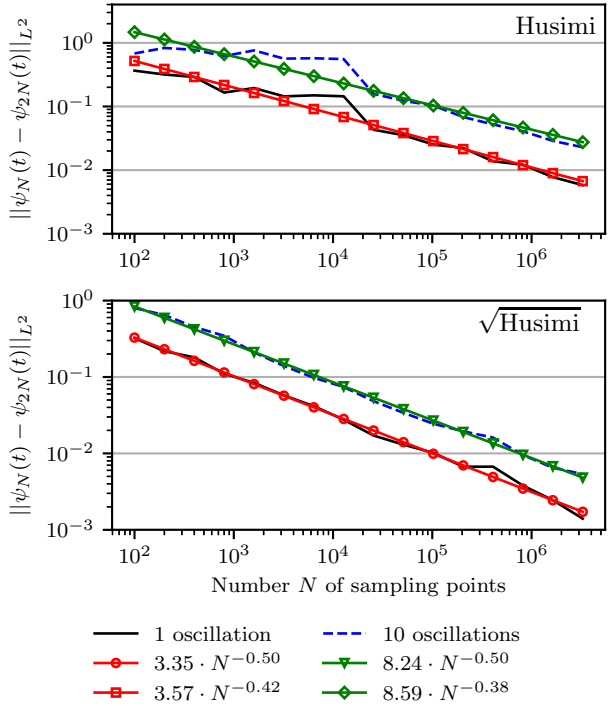}
    \caption{Same as Fig.~\ref{fig:Morse_chi_0005}, except that the anharmonicity parameter of the Morse potential was increased to $\chi=0.01$.}
    \label{fig:Morse_chi_001}
\end{figure}
\begin{figure}
 \centering
    \includegraphics{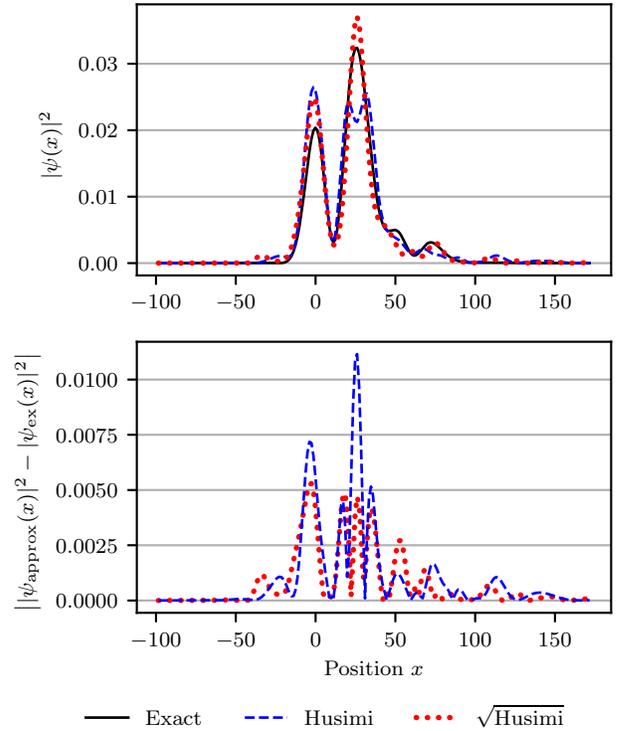}
\caption{Position probability densities (upper panel) and their absolute errors (lower panel) in a Morse potential with $\chi=0.01$ after 10 oscillations ($2020$ time steps $\approx 392$ fs). Position probability densities of the Herman--Kluk approximation for \eqref{Husimi} and \eqref{sqrt-Husimi} were computed with $N=800$ trajectories. 
}
\label{fig:Morse_001_position_1}
\end{figure}
\begin{figure}
 \centering
    \includegraphics{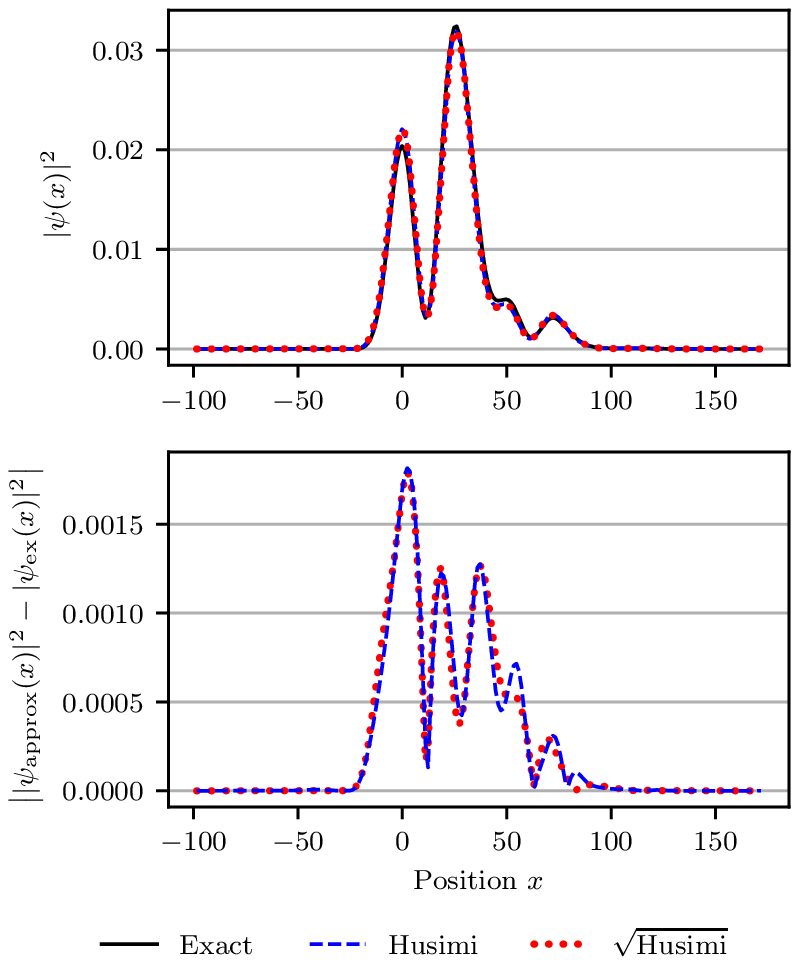}
\caption{Same as Fig.~\ref{fig:Morse_001_position_1}, except that the results were computed with $N=100\cdot2^{14}\approx 1.6 \cdot 10^6$ trajectories and, therefore, are converged.}
\label{fig:Morse_001_position_2}
\end{figure}
To investigate the convergence of the Herman--Kluk wavefunction in anharmonic systems, we consider dynamics generated by a less and more anharmonic Morse potentials. The parameters were taken from [\onlinecite{Begusic_Vanicek:2019}]. Our initial state is a Gaussian wavepacket with zero initial position and momentum, $(q_0,p_0) = (0,0)$, and with a width parameter $\gamma=0.00456$ a.u. $\approx 1000$ $\text{cm}^{-1}.$ The Morse potential
\begin{align}
    V(x)=V_{\rm eq}+D_e\left[ 1-e^{-a(x-q_{\rm eq})} \right]^2
\end{align}
is characterized by the dissociation energy $D_e$, decay parameter $a$, and the position $q_{\rm eq}$ and energy $V_{\rm eq}$ of the minimum. We considered two Morse potentials, both with $V_{\rm eq}=0.1$ and $q_{\rm eq}=20.95$ a.u., but with different values of $a$ and $D_e$. The latter two parameters, however, were chosen so that the global harmonic potential fitted to the Morse potential at $q_{\rm eq}$ had the same frequency 
\begin{align}
    \omega_{\rm eq}=\sqrt{V^{\prime\prime}(q_{\rm eq})}&=\sqrt{2D_e a^2}
    =0.0041 \text{\,a.u.} \approx 900\, \text{cm}^{-1}
\end{align}
for the two Morse potentials.
The anharmonicity of the potentials was conveniently controlled with the dimensionless parameter
\begin{align}
    \chi=\hbar \frac{\omega_{\rm eq}}{4D_e},
\end{align}
which is also reflected in the bound energy levels \cite{book_Tannor:2007}
\begin{equation}
E_n = \hbar \omega_{\rm eq}\left[\left( n+\frac{1}{2} \right) - \chi \left(n+\frac{1}{2} \right)^2\right]
\end{equation}
of a Morse oscillator.
Then $D_e$ and $a$ are given by
\begin{align}
\begin{split}
D_e&=\hbar \frac{\omega_{\rm eq}}{4\chi},
\\a&=\sqrt{2\frac{\omega_{\rm eq}\chi}{\hbar}}.
\end{split}
\end{align}
We choose two different values
\begin{align}
\chi = 0.005\quad\text{and}\quad
\chi = 0.01
\end{align}
of anharmonicity and compare the Herman--Kluk propagation with a grid-based reference quantum calculation obtained by the Fourier-split method,\cite{Feit_Fleck:1983} that is second-order accurate with respect to the time step. The position grid was set from $x=-200$ to $1500$ with $4096$ equidistant points for $\chi = 0.005$ and   
from $x=-200$ to $10000$ with $16384$ equidistant points for $\chi = 0.01$. The larger grid for $\chi = 0.01$ was required to capture oscillations of the wavefunctions in the tail region, which are due to the increased anharmonicity.
For the time propagation of the Herman--Kluk wavefunction, we used a second order St\o rmer-Verlet scheme with step size $\Delta t =8$ a.u. $\approx 0.194$ fs. 

Because the Herman--Kluk approximation is not exact in a Morse potential, to separate the statistical convergence error from the semiclassical error of the fully converged Herman--Kluk approximation, in \Cref{fig:Morse_chi_0005} we show the $L^2$-error 
\begin{align}
\norm{\psi_N(t)-\psi_{2N}(t)}
\end{align}
between the Herman--Kluk wavefunctions
calculated with $N$ and $2N$ trajectories as a function of $N$. Both wavefunctions were propagated in a Morse potential with anharmonicity $\chi=0.005$. Each panel includes the error for the fixed time $t$ after approximately one oscillation $(196$ timesteps) and ten oscillations $(1960$ timesteps). 
In addition, the convergence rates for both sampling schemes were fitted to the same power law  \eqref{Linear_Fitting}.  We observe that sampling from the square root of the Husimi density \eqref{sqrt-Husimi}, shown in the lower panel, performs better than sampling from the Husimi density \eqref{Husimi}, displayed in the upper panel.

\Cref{fig:Morse_chi_001} shows the analogous results obtained in a Morse potential with a larger anharmonicity parameter $\chi=0.01.$  Here, we display the wavefunctions after $202$ and $2020$ time steps, which are approximately the times after the first and tenth oscillations. As expected for anharmonic evolution, the error after ten oscillations is worse than after one oscillation. Increased anharmonicity also increases the error in comparison to \cref{fig:Morse_chi_0005}. 
Again, the sampling from the square root of the Husimi density \eqref{sqrt-Husimi} has consistently a lower error than \eqref{Husimi}. 

To complement the abstract convergence study of the $L^2$-error, in the upper panels of \cref{fig:Morse_001_position_1} and \cref{fig:Morse_001_position_2}, we compare the more intuitive position probability densities of the ``exact'' quantum grid-based solution with those of \eqref{Husimi} and \eqref{sqrt-Husimi}.  Both figures were obtained in a Morse potential with anharmonicity parameter $\chi=0.01$ at a time after ten  oscillations. The difference lies in the number of trajectories used. The less converged results in \cref{fig:Morse_001_position_1} were obtained with only $N=800$ trajectories, whereas the fully converged results in \cref{fig:Morse_001_position_2} employed $N=100\cdot 2^{14}$ trajectories. The lower panels of the two figures display the absolute errors of the position density for the two cases, measured with respect to the ``exact'' grid-based position density. The two panels of \cref{fig:Morse_001_position_1} confirm again that sampling from the square root of Husimi density \eqref{sqrt-Husimi} results in faster convergence than sampling from the Husimi density \eqref{sqrt-Husimi}. The upper panel of \cref{fig:Morse_001_position_2} shows that in the numerically converged regime, the Herman--Kluk propagator approximates the exact solution in this system very well, regardless whether one samples from the Husimi density or its square root. The fact that results are numerically converged is confirmed in the lower panel of \cref{fig:Morse_001_position_2}, where the errors of \eqref{Husimi} and \eqref{sqrt-Husimi} are approximately the same, which implies that the common remaining error is the error of the Herman--Kluk approximation and not the phase-space discretization error.

Analytical expressions and numerical fits to the convergence rates for all studied systems are summarized in \cref{Overview of convergence rates}.

Finally, we note that \cref{fig:Introduction} in the Introduction was based, as \cref{fig:Morse_001_position_2}, on the Morse potential with anharmonicity parameter $\chi=0.01$ and Herman--Kluk calculations used $N=100\cdot 2^{14}$ trajectories. For the computation of the position density, we used the more efficient \eqref{sqrt-Husimi}. The spectra in the upper panel were obtained by the Fourier transform of the wavepacket autocorrelation function; the Herman--Kluk autocorrelation function was computed by sampling from the Husimi density \eqref{Husimi}, because it gives the exact result (=1) at $t=0$ regardless of the number of trajectories and, therefore, converges faster at short times.

\section{Conclusion and outlook}
\begin{table}
\begin{ruledtabular}
\begin{tabular}{c c c}

&\eqref{sqrt-Husimi} &  \eqref{Husimi} \\
\hline

Initial time & &
\\
$D=1$ & $\sqrt{(4^D-1)} \cdot N^{-1/2}$ & $2.56 \cdot N^{-0.48}$ \\
$D=4$ & $\sqrt{(4^D-1)} \cdot N^{-1/2}$ & $19.3 \cdot N^{-0.36}$ \\
&&\\

Harmonic pot.  & $\mathbb{V}\big[\ket{\tilde{\psi}(t)}\big]^{1/2}\cdot N^{-1/2}$ &  \\
&  with variance &
\\ & from \eqref{VarianceHarmonicOverTime} & \\ 
&&\\
Morse pot.: $\chi=0.005$  && \\

1 oscillation & $2.19 \cdot N^{-0.49}$ & $2.14 \cdot N^{-0.41}$ \\
10 oscillations & $4.68 \cdot N^{-0.51}$ & $3.24 \cdot N^{-0.36}$ \\
&& \\
Morse pot.: $\chi=0.01$ && \\
1 oscillation & $3.35 \cdot N^{-0.50}$ & $3.57 \cdot N^{-0.42}$ \\
10 oscillations & $8.24 \cdot N^{-0.50}$ & $8.59 \cdot N^{-0.38}$ \\

\end{tabular}
\end{ruledtabular}
\caption{Summary of convergence rates for \eqref{sqrt-Husimi} and \eqref{Husimi} at initial time, in a harmonic oscillator and two Morse potentials after one and ten oscillations.}
\label{Overview of convergence rates}
\end{table}


We compared two different sampling strategies for evaluating the semiclassical wavefunction evolved with the Herman--Kluk propagator. For the initial phase-space sampling, we either used the Husimi density \eqref{Husimi} or its square root \eqref{sqrt-Husimi}. We showed that the square root sampling produces a Monte Carlo integrand with finite second 
moment, while the Husimi sampling comes with an undesirable infinite second moment. The numerical experiments 
for the harmonic oscillator and two Morse oscillators with different extents of anharmonicity confirm that the infinite second moment results in a slower convergence of the Monte Carlo estimator. 
Therefore, we recommend the square root approach \eqref{sqrt-Husimi} whenever the Herman--Kluk propagator is used directly for approximating the quantum-mechanical wavefunction and the $L^2$-error 
of the wavefunction approximation is the relevant accuracy measure. A follow-up paper applying a similar analysis to the autocorrelation function as well as to the expectation values computed with the Herman--Kluk propagator is in preparation.

\section*{Acknowledgments}
The authors acknowledge financial support from the European Research Council (ERC) under the European Union’s Horizon 2020 Research and Innovation Programme (Grant Agreement No. 683069–MOLEQULE) and from the EPFL.

\appendix

\section{Boundedness of Herman--Kluk prefactor} \label{Appendix:1}
For notational simplicity we consider the case $\gamma = \mathrm{Id}_{D}.$ A general $\gamma$ does not change the argument, but it requires to exchange $M$ by 
\begin{align}
    \begin{pmatrix}
    \mathrm{Id}_D & 0 \\ 0 & \gamma^{-1}
    \end{pmatrix} \cdot M \cdot \begin{pmatrix}
    \mathrm{Id}_D & 0 \\ 0 & \gamma
    \end{pmatrix}
\end{align}
in the following. We recall that 
\begin{align}
\begin{split}
&R(t,z)
\\&=\sqrt{2^{-D} \det{\left(M_{qq}+ M_{pp}  - i M_{qp} +i M_{pq}\right)}}
\\&=: \sqrt{\det{F}}
\end{split}
\end{align}
with
\begin{align}
F = (\mathrm{Id}_{2D} + M) - iJ \cdot (\mathrm{Id}_{2D} - M).
\end{align}
Following [\onlinecite[Lemma~5.6]{Lasser_Lubich:2020}], for $F$ one calculates
\begin{align}
F^\dagger \cdot F = 2(\mathrm{Id}_{2D} + M^T\cdot M).
\end{align}
Hence, 
\begin{align}
\begin{split}
|R(t,z)|^2 &= \sqrt{\det(F^\dagger \cdot F)} 
\\& = \sqrt{2^{D}\det{(\mathrm{Id}_{2D} + M^T\cdot M)}}.
\end{split}
\end{align}
With this representation, we can link growth and boundedness of $R(t,z)$ directly with growth and boundedness of the stability matrix, and it is the largest eigenvalue of the Hessian $\textrm{Hess} \, h$ that ultimately gives upper bounds on $R(t,z)$.


\section{Number of trajectories needed for convergence} \label{Appendix:2}
In the following, we give a lower bound as well as an asymptotic estimate for the number of trajectories $N$ needed ``for convergence,'' or, more precisely, needed to guarantee that the probability of the convergence error exceeding a specified threshold $\epsilon$ is below a prescribed value $p$. The following analysis holds for any potential $V$ in any number of dimensions $D$, but assumes that the variance $\sigma^2=\mathbb{V}[\ket{\psi(t)}]$ of the estimator is finite.

On one hand, Chebyshev's inequality\cite{durrett:2019} 
\begin{align}
        \mathbb{P}\left(\norm{\psi_N(t) - \psi^{\rm HK}(t)} \geq \epsilon \right) \leq  \frac{\sigma^2}{N \epsilon^2}
\end{align}
implies that the number of trajectories $N$ must obey the inequality
\begin{align}
        N\geq \frac{\sigma^2}{p \epsilon^2}
\end{align} 
in order that the probability of the convergence error $\norm{\psi_N(t) - \psi^{\rm HK}(t)}$ exceeding $\epsilon$ be at most $p$.


On the other hand, the central limit theorem\cite{durrett:2019} states that the random variable
\begin{align}
    Z_N:= \frac{\sqrt{N}}{\sigma} \left( \psi_N(t) -\psi^{\rm HK}(t)  \right)
\end{align}
converges in distribution to the standard normally distributed random variable. Hence, the probability for the convergence error to be above this threshold is 
\begin{align}
    \begin{split}
        &\mathbb{P}\left(  \norm{\psi_N(t) - \psi^{\rm HK}(t)} \geq \epsilon \right)
        \\& \approx (2\pi)^{-1/2} \int_{\frac{\sqrt{N}}{\sigma} \epsilon}^{\infty} e^{-t^2/2} \, dt.
    \end{split}
\end{align}
For a given probability $p$ that $\epsilon$ is exceeded, this leads to the asymptotic behaviour
\begin{align}
    N \approx \frac{\sigma^2}{2\epsilon^2}\left[\operatorname{erfc}^{-1}(2p)\right]^2,
\end{align}
where $\operatorname{erfc}$ is the complementary error function.\cite{Dyke:2013}

\section*{References}
\bibliographystyle{apsrev4-2}
\bibliography{addition_HK}

\end{document}